# Sign change of the anomalous Hall effect and the anomalous Nernst effect in Weyl semimetal CeAlSi

Md Shahin Alam [1,*], Amar Fakhredine [2], Mujeeb Ahmed [1], P.K. Tanwar [1], Hung-Yu Yang [3], Fazel Tafti [3], Giuseppe Cuono [1], Rajibul Islam [1], Bahadur Singh [4], Artem Lynnyk [2], Carmine Autieri [1,†], Marcin Matusiak [1,5,‡]

1. *International Research Centre MagTop, Institute of Physics, Polish Academy of Sciences, Aleja Lotników 32/46, PL-02668 Warsaw, Poland*
2. *Institute of Physics, Polish Academy of Sciences, Aleja Lotników 32/46, PL-02668 Warsaw, Poland*
3. *Department of Physics, Boston College, Chestnut Hill, Massachusetts 02467, USA*
4. *Department of Condensed Matter Physics and Materials Science, Tata Institute of Fundamental Research, Mumbai 400005, India*
5. *Institute of Low Temperature and Structure Research, Polish Academy of Sciences, ul. Okólna 2, 50-422 Wrocław, Poland*

We report the anomalous Hall effect (AHE) and the anomalous Nernst effect (ANE) data for the non-collinear Weyl semimetal CeAlSi. The anomalous Hall conductivity ($\sigma_{ij}^A$) was measured for two different orientations of the magnetic field (*B*), namely $\sigma_{yz}^A$ for *B* || *a* and $\sigma_{xy}^A$ for *B* || *c*, where *a* and *c* denote the crystallographic axes. We find that $\sigma_{xy}^A$ and $\sigma_{yz}^A$ are of opposite sign and both are large below the Curie temperature ($T_C$). In the paramagnetic phase, $\sigma_{xy}^A$ raises even more and goes through a maximum at $T \approx 170$ K, whereas the absolute value of $\sigma_{yz}^A$ decreases with increasing temperature. The origin of the sign difference between $\sigma_{xy}^A$ and $\sigma_{yz}^A$ was attributed to the reconstruction of the band structure under the variation of the spin orientation. Further, in a system where humps in the AHE are present and scalar spin chirality is zero, we show that the **k**-space topology plays an important role to determine the transport properties at both low and high temperatures. We also observed the anomalous contribution in the Nernst conductivity ($\alpha_{xy}^A$) measured for *B* || *c*. $\alpha_{xy}^A/T$ turns out to be sizeable in the magnetic phase and above $T_C$ slowly decreases with temperature. We were able to recreate the temperature dependences of $\sigma_{xy}^A$ and $\alpha_{xy}^A/T$ in the paramagnetic phase using a single band toy-model assuming a non-zero Berry curvature in the vicinity of the Weyl node. A decisive factor appears to be a small energy distance between the Fermi level and a Weyl point.

__________
[*] shahin@magtop.ifpan.edu.pl, [†] autieri@magtop.ifpan.edu.pl, [‡] matusiak@magtop.ifpan.edu.pl

# 1. Introduction

Topological Weyl semimetals (WSMs) are characterised by linear energy dispersions of the valence and conduction bands touching each other in momentum space at Weyl nodes [1–4]. The emergence of massless Weyl fermions as low-energy excitations manifests in many exotic physical effects like presence of Fermi arcs on the surface [5], chiral anomaly induced negative magnetoresistance [6], chiral zero sound effect [7,8] etc. The subclass of WSMs that also exhibit magnetic ordering is a particularly interesting object of study [9–15]. These materials allow, for example, the manipulation of the anomalous Hall and anomalous Nernst effect [16,17], which is interesting from both scientific and applicative points of view. Recently, huge efforts have been made to investigate the sign change of the anomalous Hall effect (AHE) in Weyl fermions or closely relevant systems as the collinear ferromagnet $SrRuO_3$ [18]. It turns out that many factors, such as the value of the magnetization [19], the presence of the interface which can tune the spin-orbit coupling (SOC) or breaking of the inversion symmetry [20], can change the sign of AHE. Moreover, in the presence of the sign change, the anomalous Hall effect may take values smaller than other features, such as humps in the hysteresis loop of AHE [19,20]. The presence of these humps seems to be particularly favoured by a large spin-orbit coupling as well as the absence of inversion symmetry [19,20] and in CeAlSi they were dubbed the loop Hall effect [21]. In CeAlSi, the humps are related to the **k**-space topology [21]. In order to get more insight into the **k**-space topology that governs these humps in CeAlSi, we investigated CeAlSi focusing on the transport properties in magnetic fields and looking for the sign change of the AHE. Additionally, we also measured the anomalous Nernst effect (ANE), which response to non-zero Berry curvature around the Fermi energy is different than that expected for AHE. In the paramagnetic phase of CeAlSi, the simultaneous temperature evolution of both ANE and AHE can be well described by a simple model assuming presence of the Weyl node about 20 meV from the Fermi level. The paper is organized as follows: we describe the material and methods in Section 2; in Section 3, we present our experimental results; in Section 4, we discuss our results and in Section 5 we summarize our conclusions.

**2. Material and methods**

CeAlSi single crystals were grown by a self-flux method using the Canfield crucible sets. The starting materials were weighed in the ratio Ce:Al:Si = 1:10:1, placed inside a crucible in an evacuated quartz tube, heated to 1000 °C at 3 °C/min, held at 1000 °C for 12 h, cooled to 700 °C at 0.1 °C/min, stayed at 700 °C for 12 h, and centrifuged to decant the residual Al flux.

For electrical and thermo-electrical transport measurements a suitable single crystal was cut into a plate with dimensions of 1.4 x 1.3 x 0.4 mm$^3$ with sides parallel to the natural crystallographic *a* x *a* x *c* – axes. The electrical and thermal currents were applied along the longest side of the sample (*a*- axis), while the magnetic field (*B*) was applied parallel to the *c*- axis and perpendicular to the thermal and electrical currents. For electrical measurements with *B* applied along the crystallographic *a* – axis we selected another single crystal, which was cut into a plate with dimensions of 0.2 x 2.9 x 0.6 mm$^3$ (*a* x *a* x *c*). The electrical current (*J*) was applied along *a* – axis and *B* was perpendicular to the current and parallel to another *a* – axis.

For the electrical measurements, the contacts were arranged in the Hall-bar geometry with 25 μm thick gold wires attached to the sample using DuPont 4929N silver paste. During the measurements of thermoelectric power ($S_{xx}$) and Nernst signal ($S_{yx}$) sample was mounted between two blocks made of phosphor bronze. The temperature difference was determined using two Cernox thermometers and the thermal gradient was implied using a strain gauge as a resistive heater. For selected temperatures, the magnetic field was swept from -14.5 to +14.5 T to extract the field voltage components that were anti-symmetric and symmetric in *B*. All the presented data were symmetrized and anti-symmetrized for the positive and negative magnetic fields.

Magnetic properties of the sample have been investigated using Quantum Design Magnetic Property Measurement System MPMS XL equipped with a superconducting quantum interference device. The reciprocating sample option has been chosen to provide a precision of about $10^{-8}$ emu during the direct current (dc) measurements. A magnetic moment as a function of external dc magnetic field has been measured in the range of– 7 to +7 T after cooling at zero field. To study the temperature dependences of susceptibility, alternative current (ac) option has been utilized. AC field of 1 Oe amplitude and 1 kHz of frequency have been applied during the measurements.

## 3. Results

The electrical transport properties of CeAlSi were studied for two different orientations of the magnetic field (*B*), which was applied along the *a*-axis (magnetically easy) or along the *c*-axis (magnetically hard). The temperature (*T*) dependence of the longitudinal resistivity ($\rho_{xx}$) for the electrical current (*J*) parallel to the *a*-axis and *B* = 0 T (see Fig. 1) shows semimetallic behaviour with the residual resistivity ratio of 3.2 (see Fig. 1). A kink at $T_C \approx 8.5$ K appears due to the transition from the high-temperature-paramagnetic (PM) phase to the low-temperature ferromagnetic (FM) phase [21]. The value of the transition temperature and overall temperature dependence of $\rho_{xx}$ agree well with the previous reports [21–23]. The magnetic field dependences of the longitudinal resistivity ($\rho_{xx}$) and Hall resistivity ($\rho_{yx}$) (*B* ∥ *c* and *J* ∥ *a*), as well as $\rho_{yy}$ and $\rho_{zy}$ (*B* and *J* ∥ *a*, *B* ⊥ *J*), are shown in Fig. 2. The Hall resistivities for both orientations of *B* (Fig. 2(b,d)), becomes negative at low temperature and high magnetic field, which indicates that the electron-like charge carriers dominate electrical transport in this region. Specifically, at *B* =

14.5 T , $\rho_{yx} \approx$ - 35 µΩ cm for $T \approx$ 5.4 K and the $\rho_{zy} \approx$ - 65 µΩ cm for $T \approx$ 1.7 K. However, the high-field slope of the $\rho_{yx}$ Hall resistivity evolves with temperature and becomes slightly positive at room temperature (d$\rho_{yx}$/d$B$ = 1.2 × $10^{-9}$ m$^{-3}$ C$^{-1}$ at $T$ = 301 K), which might be due to the slowly increasing with temperature role of holes in the electrical transport.

A prominent characteristic of the Hall resistivity in CeAlSi is its nonlinear field dependence (see Fig. 2(b)). Although this type of behavior might be due to simultaneous contributions from different types of charge carriers [24,25], it seems to be not the case in CeAlSi. In fact, $\rho_{yx}$ varies linearly at high-field, but $\rho_{yx}(B)$ does not extrapolate to $\rho_{yx}$ = 0 µΩ cm at $B$ = 0 T. The field dependence of $\rho_{yx}$ cannot be satisfactorily explained within the two-band model approach as shown in Fig. S1 in the Supplemental Materials (SM) [26] (see also references therein [27–31] therein) , which is in line with the previous reports indicating that the transport in CeAlSi is dominated at low temperatures by a single-type of charge carriers [32]. A likely cause for a non-linear behavior of the Hall resistivity in CeAlSi is a contribution from the anomalous Hall effect (AHE) [21,33–36]. In general, the magnetic field dependences of Hall resistivity in presence of the anomalous contribution can be expressed as:

$$\rho_{yx} = R_0 B + \rho_{yx}^A ,  \qquad (1)$$

where $R_0$ is the ordinary Hall coefficient and $\rho_{yx}^A$ is anomalous Hall resistivity. To determine $R_0$ and to separate $\rho_{yx}^A$ (or $\rho_{zy}^A$) from the total Hall resistivity, $\rho_{yx}(B)$ (or $\rho_{zy}(B)$) was fitted with a linear function in the high field regime (> 3 T) (see Fig. 3(a,b)). At low temperature, the fitting range of the former was restricted to $B_{\max}$ = 9 T, because of a change in $\rho_{yx}(B)$ slope happening at this field. This anomaly is also visible in the field dependences of $\rho_{xx}$ (Fig. 2(a)) and it is likely related to the magnetic phase transition caused by increasing magnetic field perpendicular to the initially FM ordered spins [37]. The extracted field dependences of the anomalous Hall resistivity for $B \parallel c$ ($\rho_{yx}^A$) and $B \parallel a$ ($\rho_{zy}^A$) are presented in Fig. 3(c) & (d). In the FM phase, both $\rho_{yx}^A$ and $\rho_{zy}^A$ are sizeable, but different in sign and magnitude. In the PM phase, the absolute value of $\rho_{yx}^A$ becomes significantly larger than $\rho_{zy}^A$.

In general, the AHE can be of intrinsic or extrinsic origin and the latter can be due to the skew-scattering or side-jump processes [27]. If there were a contribution from extrinsic mechanism, one would expect a specific relation between the resulting anomalous Hall conductivity (AHC, $\sigma_{ij}^A$) and the longitudinal conductivity ($\sigma_{ii}$). Namely, for the skew-scattering, the AHC should follow a linear relationship with the longitudinal conductivity, whereas for the side-jump (which is expected to be somewhat smaller [38]) $\sigma_{ij}^A \sim \sigma_{ii}^2$ [39]. In our data we observe neither of them (Fig. S2 in SM [26]). Furthermore, in topological semimetals with the Fermi level in the vicinity of Weyl nodes, the AHE has been predicted to be predominantly intrinsic and determined by the location of the Weyl points [40,41]. Since CeAlSi has been identified as a Weyl semimetal with Weyl nodes close to the Fermi level [21], one can expect a nonzero AHE due to its non-trivial topological properties [21–23,42]. These are expected to manifest themselves also in other transport phenomena.

The magnetic field dependence of thermoelectric power with the thermal gradient $\nabla T \parallel a$ and $B \parallel c$, are presented in Fig. 4(a). $S_{xx}(B)$ can be satisfactorily fitted with the semi-classical phenomenological model proposed by Liang et al., [43]:

$$S_{xx}(B) = S_{xx}^0 \frac{1}{1 + (\mu B)^2} + S_{xx}^\infty \frac{(\mu B)^2}{1 + (\mu B)^2}, \qquad (2)$$

where $S_{xx}^0$ and $S_{xx}^\infty$ are the amplitudes of the thermopower at zero and high field limits respectively, and $\mu$ is the mobility of charge carriers. A shift of $S_{xx}^0$ from negative at low temperatures to positive at high temperatures can be a sign of an increase in the participation of holes, which is consistent with the Hall resistivity data discussed earlier. If these holes have low mobility, then their contribution will be only slightly field dependent and will not disturb the fitting procedure, which in fact works well for all the data (see Fig. 4(a)). At low temperatures, we restricted again the fitting field range due to the change in $S_{xx}(B)$ slope at $B \sim 8 - 9$ T owing to the aforementioned field-induced transition. The temperature dependence of $S_{xx}$ at zero magnetic field is shown in Fig. 1.

Figure 4 presents the field dependences of the Nernst effect signal measured for a configuration analogous to $\rho_{yx}$, i.e. $\nabla T \parallel a$ and $B \parallel c$. Results are fitted with the empirical model [44] describing the behaviour of the Nernst effect in a topologically non-trivial material. Here, the total Nernst signal is similar to the Hall resistivity and is divided into a normal ($S_{yx}^N$) and an anomalous ($S_{yx}^A$) part:

$$S_{yx} = S_{yx}^N + S_{yx}^A, \qquad (3)$$

Where their field dependences are expressed as:

$$S_{yx}^N = S_0^N \frac{\mu}{1+(\mu B)^2}, \qquad (4)$$

$$S_{yx}^A = S_{yx}^A \tanh(B/B_s), \qquad (5)$$

$\mu$ is the mobility, and $B_s$ is the saturation field at which the plateau of the anomalous signal is reached. Apparently, the field dependences of $S_{yx}$ cannot be described only by the conventional Nernst contributions (Eqn. 4) (see Fig. S1 in SM [26] ), but they are very well approximated when the anomalous component (Eqn. 5) is taken into consideration (see Fig. 4b). The temperature dependence of the normalized (divided by temperature) $S_{yx}^A$ is displayed in the inset of Fig. 5b. In the FM phase $S_{yx}^A/T$ steeply increases with decreasing the temperature (reaching $S_{yx}^A \approx -0.1\ \mu V K^{-2}$ at 2.5 K), but also in the PM phase there is a non-vanishing anomalous contribution (in order of $S_{yx}^A \approx -0.02\ \mu V K^{-2}$) slowly decreasing with temperature.

## 4. Discussion

The anomalous Hall ($\sigma_{ij}^A$) and transverse thermoelectric ($\alpha_{ij}^A$) conductivities can be calculated as [45–47]:

$$\sigma_{ij}^A = \frac{\rho_{ji}^A}{\rho_{ii}^2}, \tag{6}$$

$$\alpha_{ij}^A = S_{ji}^A \sigma_{ii} - S_{ii}\sigma_{ij}^A, \tag{7}$$

if $\rho_{ii} \gg \rho_{ji}^A$. The resulting temperature dependences of $\sigma_{xy}^A(T)$ and $\alpha_{xy}^A/T\,(T)$ are presented in Fig. 5. Two temperature regions can be distinguished: (i) $T < T_C$ (FM phase) and (ii) $T > T_C$ (PM phase).

(i) In the ferromagnetic phase $\sigma_{xy}^A$ gradually increases with decreasing temperature reaching $\sim 550\ \Omega^{-1}cm^{-1}$ at $T = 5.4$ K (Fig. 5a). The loop Hall effect (LHE) was reported to occur in CeAlSi in the FM phase with $B \parallel c$ field orientation [21], but the appearance of this phenomenon changes from sample to sample depending on a slight off-stoichiometry of Si and Al [21]. In this material class, the Weyl nodes are generated due to the lack of inversion symmetry in the non collinear phase [21,22], while for the ferromagnetic phase Weyl points can be generated also by the breaking of time-reversal symmetry [42]. A recent study on CeAlSi suggested a nontrivial $\pi$ Berry phase that has been experimentally reported in the FM regime for the magnetic field oriented along the $c$-axis [23].

Similarly to $\sigma_{xy}^A$, we also determined the $T$-dependence of $\sigma_{yz}^A$ for the magnetic field oriented along the easy axis. $\sigma_{yz}^A(T)$ is presented in the inset of Fig. 5a. We found $\sigma_{yz}^A \approx -380\ \Omega^{-1}cm^{-1}$ at $T = 1.7$ K, a magnitude that is consistent with the previous reports [21]. Differences in values of $\sigma_{xy}^A$ and $\sigma_{yz}^A$ can be attributed to the anisotropic electronic structure of CeAlSi, while the observed sign change may be relevant for the detection of topological features in the AHE. Its occurrence, for example, was recently associated with the presence of hump-like features in $\rho_{yx}(B)$ [19,48]. A physical origin of this anomaly is under strong debate, but it could derive from topological effects in the k-space and/or in the real space. In CeAlSi appearance of the analogous loop Hall effect appears to dependent on the position of the Fermi level [21].

We study the magnetic configurations with spins along the *a*- and *c*-axis (*x* and *z*, respectively) in addition to the non-collinear magnetic configuration that is the ground state. Using density functional theory (DFT) and wannierization techniques, we perform the self-consistent and band structure calculations for different magnetic configurations to investigate the sign change of the AHE in the magnetic phase below $T_C$ = 8.5 K.

From the self-consistent calculation, we note that the magnetization is mostly coming from the 4f-electrons of Ce. The local magnetic moment per Ce atom is approximatively constant in all magnetic configurations. The magnetic moment for the 4f-orbitals is 0.85-0.89 $\mu_B$ where the lowest value is for the non-collinear magnetic configuration and the highest for both collinear configurations. We have an intrinsic magnetic moment from 4f-orbitals and an induced magnetic moment on the 5d-orbitals of Ce that is 0.03 $\mu_B$ within DFT. The f-electrons induce a ferromagnetic moment on the d-electrons of Ce in the same fashion as happens in the $EuTiO_3/SrTiO_3$ system [49]. The presence of magnetic f-electrons far from the Fermi level and d-electrons at the Fermi level makes difficult to produce a simplified tight-binding model containing both d- and f-orbitals. The bands associated with the Weyl points mainly come from the d-electrons of Ce, sp-electrons of Al and Si as clearly described in the local density of states (see part E in SM [26]).

We report in Fig. 6(a,b,c) the magnetic configurations with the Ce spins along the *a*-axis, *c*-axis and with the non-collinear configuration, respectively, where *a* and *c* are the lattice constants of the conventional unit cell shown in the figure. The band structures associated with these magnetic configurations are in the respective bottom panels in Fig. 6(d,e,f). The main features of the three band structures are the same, but the different magnetic configurations slightly move the details of the low energy features and switch the position of the Weyl points [50]. One relevant change for the AHE appears along the high-symmetry path **Σ-N** and N-**Σ$_i$** where we can see at the position of the vertical arrows that the bands close to the Fermi level are slightly lower in energy in the case of the magnetic configuration with spin along the *c*-axis shown in Fig. 6(e), as a consequence the minimum of the conduction band along **ΓX** goes higher in energy in Fig. 6(e). Therefore, the AHE will be modified by an energy shift approximatively equal to the difference between the Weyl points for the case with spin along the *c*-axis ($E_{WP}^z$) and the *a*-axis ($E_{WP}^x$). Defined as $\Delta E_{WP} = E_{WP}^z - E_{WP}^x > 0$, this shift will reflect in the AHE calculations. Basically, the different magnetic orderings influence the position of the Fermi level and the energy position of the Weyl points, and the anticrossing points close to the Fermi level.

It is known that close to the high-symmetry line **ΓX** there are several Weyl points [42]. In CeAlSi, there are Weyl points from the breaking of the inversion symmetry and Weyl points from the breaking of the time-reversal symmetry. The Weyl points from the breaking of the time reversal present along **ΓX** are expected to be more sensitive to the orientation of the magnetic order, therefore strong changes in the AHE are expected.

Given the three band structures in Fig. 6(d,e,f), we extracted the Wannier tight-binding model (see SM for details [26]) and calculated the anomalous Hall effect for the three magnetic configurations shown in Fig. 7. We report $\sigma_{xy}$ for the magnetic configuration with spin along *c*-axis (hard-axis), and $\sigma_{yz}$ for the configurations with spins along the *a*-axis (easy-axis) and the non-collinear phase. In the calculated energy range between -0.5 and 0.5 eV, the calculated AHC is always positive except for a negative spike present for all configurations. While for in-plane magnetic configuration this spike is at the Fermi level, for the out-of-plane magnetic configuration this negative spike is shifted by the quantity $\Delta E_{WP}$ deriving from the band structure effects. This implies that the change of the magnetization from the *a*- to *c*-axis plays a role in inverting the sign of the anomalous Hall conductivity. The AHC is positive for the out-of-plane magnetic field and negative for the in-plane magnetic field in agreement with the experimental results reported in Fig. 5a. The presence of consecutive and negative large values of the Berry curvature is a signature of the Weyl points, indeed, in a simplified Weyl points model, the Berry curvature goes from being strongly positive to strongly negative when you go from below to above the Weyl points [51] (see part D of SM [26]). Hence, the sign change of AHE comes directly from the presence of the **k**-space topology (Weyl points) close to the Fermi level.

Our theoretical results developed at low temperatures could be qualitatively valid also above $T_C$, where the magnetization rotates from the easy axis towards the axis of the applied strong magnetic field. Since the Weyl points at the Fermi level do not come from 4f-electron bands, we expect that AHE is weakly dependent if the induced magnetic moment on 5d-electrons comes from the 4f-Ce intrinsic magnetic moment or from the external magnetic field. Therefore, the AHE above Curie temperature can be large too and AHE below and above $T_C$ can be of the same order of magnitude. Therefore, the large AHE in the paramagnetic phase emerges due to the presence of the **k**-space topology (Weyl points) close to the Fermi level. Indeed, the Weyl points are close to the Fermi level giving a large contribution even in presence of the external magnetic field.

(ii) In the paramagnetic phase, the anomalous Hall conductivity for $B \parallel a$ ($\sigma_{yz}^A$) decreases with increasing temperature and practically vanishes at room temperature. On the contrary, the anomalous Hall conductivity for $B \parallel c$ ($\sigma_{xy}^A$) goes through a maximum at $T \approx 170$ K (see Fig. 5a), and reaches higher values than in the FM phase. The corresponding anomalous Nernst conductivity (ANC, $\alpha_{xy}^A/T$) slowly decreases with increasing temperature (see Fig. 5b).

It is worth noting here that a sizeable anomalous response was already reported in other non-magnetic topological materials [52,53]. The lack of correlation between magnetization and the ANE was even used to indicate that the observed phenomenon is due to non-zero Berry curvature [44]. In topological semimetals the AHE as well as the ANE originate from large Berry curvature generated by Weyl nodes and their presence in the paramagnetic phase of CeAlSi was recently confirmed experimentally [22]. In the presence of a finite Berry curvature, $\sigma_{xy}^A$ and $\alpha_{xy}^A$ can be calculated as [53,54]:

$$\sigma_{xy}^A = \frac{e^2}{\hbar} \sum_n \int \frac{d^3k}{(2\pi)^3} \Omega_{xy}^n f_n, \quad (8)$$

$$\alpha_{xy}^A = -\frac{1}{T}\frac{e}{\hbar} \sum_n \int \frac{d^3k}{(2\pi)^3} \Omega_{xy}^n \left[(E_n - E_F)f_n + k_B T \ln\left(1 + exp\frac{(E_n-E_F)}{-k_B T}\right)\right], \quad (9)$$

where $f_n$ is the Fermi-Dirac distribution, $E_F$ represents the Fermi energy, $\Omega_{xy}^n$ is the Berry curvature and $E_n$ are the eigenenergies for eigenstates $n$. From the above equations, a general form of ANC and AHC can be written as [51]:

$$\lambda_{xy} = \frac{e^2}{\hbar} \sum_n \int \frac{d^3k}{(2\pi)^3} \Omega_{xy}^n w_\lambda (E_n - E_F) \text{ with } \lambda = \sigma, \alpha. \quad (10)$$

Hence, both anomalous conductivities are basically the product of $\Omega_{xy}^n$ and weighting factor ($w$), where the latter reads as [51]:

$$w_\sigma(E_n - E_F) = f_n^T(E_n - E_F), \quad (11)$$

$$w_\alpha(E_n - E_F) = -\frac{1}{eT}\left[(E_n - E_F)f_n^T + k_B T \ln\left(1 + exp\frac{(E_n-E_F)}{-k_B T}\right)\right], \quad (12)$$

Here $f_n^T$ is the Fermi - Dirac distribution function at a given temperature. To model the temperature dependences of $\sigma_{xy}^A$ and $\alpha_{xy}^A$ in the PM phase, we introduce a single-band toy model including a non-zero Berry curvature in the vicinity of the Weyl points. $\Omega_{xy}(E)$ is simplistically assumed to change linearly at the Weyl node from positive to negative (see in Fig. S4 in SM [26]). To match the experimental results, we restricted the energy range of non-zero $\Omega_{xy}$ to $E_F \pm 25$ meV, which is similar to the range reported in Ref [51]. As for the energy distance between the Fermi level and a Weyl node, the electronic structure calculations reported by [21] for CeAlSi indicate two sets $W_1$ and $W_2$ present close to $E_F$, which are expected to dominate the low energy physics of this material [21]. Each set contains different Weyl points defined as $W_1^{1,2,3,4}$ and $W_2^{1,2,3,4}$. In our model, the Weyl node is placed at -20 meV away from the $E_F$, which is consistent with the position of the $W_2$ nodes [21]. The calculated energy dependences of AHC, ANC, and $w$ at room temperature are shown in Fig. S4 of the SM [26]. The temperature dependences of $\sigma_{xy}^A$ and $\alpha_{xy}^A/T$ calculated using Eqns. 8 & 9 are presented in Fig. 5 as solid lines along with the experimental data. They appear to be governed by a broadening of the Fermi function with temperature, which allows states further away from the $E_F$ to be included in the integration [44,52,53].

Apparently, this crude approach reproduces the characteristics of the experimental data quite well, reflecting differences between energy dependences of weighting factors for the anomalous Hall and Nernst effects. Namely, the calculated $\sigma_{xy}^A$ increases up to $T \approx 170$ K, reaches a maximum, and then decreases, while $\alpha_{xy}^A/T$ slowly decreases with the increasing temperature at a rate similar to the one observed in the experiment. Moreover, the signs of both $\sigma_{xy}^A$ and $\alpha_{xy}^A/T$ also match the experimental data.

## 5. Conclusion

We studied the anomalous Hall and Nernst effects in the non-collinear Weyl semimetal CeAlSi from room to low temperature. In the ferromagnetic phase, the anomalous Hall conductivity turns out to be positive for the magnetic field applied along the magnetically hard axis ($\sigma_{xy}^A > 0$) and negative for *B* parallel to the easy axis ($\sigma_{yz}^A < 0$). Density functional theory calculations attributed the different signs of the AHE to a shift of Weyl points along the Γ-X direction and this shift is induced by the reconstructions in the band structure driven by the magnetic configuration. In the paramagnetic phase, $\sigma_{yz}^A$ significantly decreases, whereas $\sigma_{xy}^A$ reaches values even higher than at the low-temperature limit. The temperature dependence of $\sigma_{xy}^A$ as well as the respective Nernst conductivity ($\alpha_{xy}^A$) can be well approximated using a simple model assuming the presence of a Weyl point in the vicinity of the Fermi level. Properties of the anomalous Hall and Nernst effects appear to be dominated by **k**-space topology at both low and high temperatures.

**Note added in proof**: After submission of this paper, we became aware of a very recent work, which also reports on the anomalous Hall and Nernst effect in CeAlSi [55].

## 6. Acknowledgments

We acknowledge W. Brzezicki, R. Citro, and K. Das for the useful discussions. This work is supported by the Foundation for Polish Science through the International Research Agendas program co-financed by the European Union within the Smart Growth Operational Programme. A.F. was supported by the Polish National Science Centre under project no. 2020/37/B/ST5/02299. We also acknowledge the access to the computing facilities of the Interdisciplinary Center of Modeling at the University of Warsaw, Grant G84-0, GB84-1 and GB84-7. We acknowledge the CINECA award under the ISCRA initiative IsC93 "RATIO" and IsC99 "SILENTS" grants, for the availability of high-performance computing resources and support. The work at Boston College was funded by the National Science Foundation under award number DMR-2203512.

**Figures**

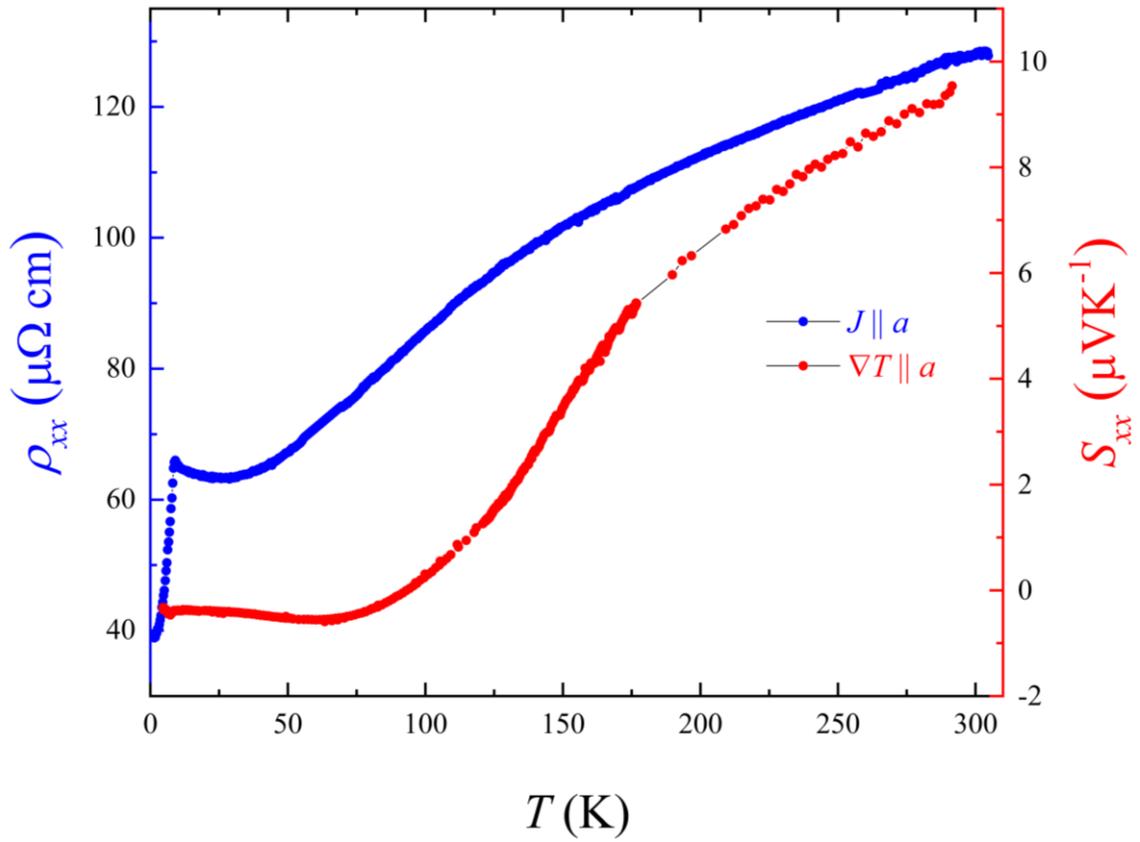

**Figure 1**. (Color online) Temperature dependences of the zero field resistivity ($\rho_{xx}$) and the thermoelectric power ($S_{xx}$) of the non-collinear Weyl semimetal CeAlSi with the current ($J$) or thermal gradient ($\nabla T$) applied parallel to $a$ – axis.

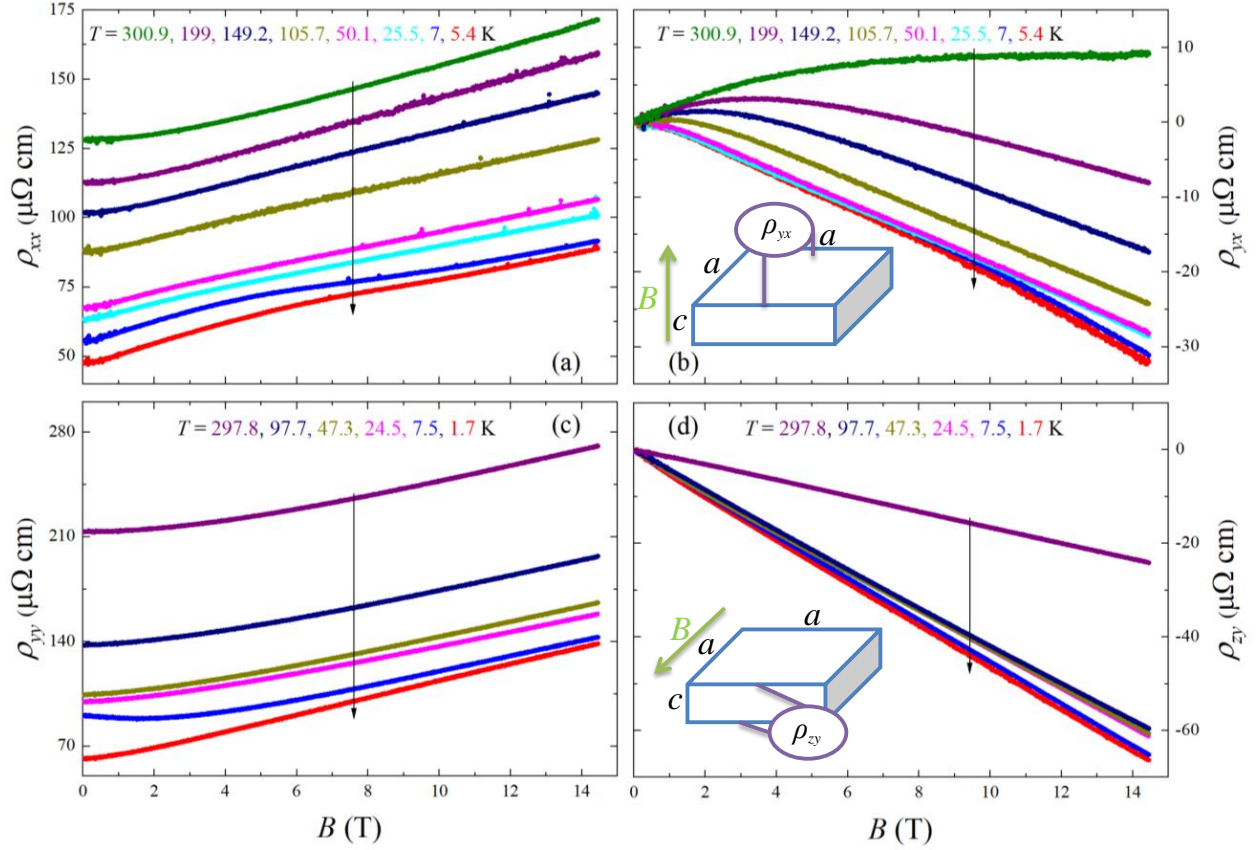

**Figure 2.** (Color online) Magnetic field dependences of the longitudinal resistivity and Hall resistivity in CeAlSi for two different configuration of the sample. For $J \parallel a$ and $B \parallel c$, (a) Longitudinal resistivity ($\rho_{xx}$); (b) Hall resistivity ($\rho_{yx}$) and $J \parallel a$ and $B \parallel a$ and $B \perp J$; (c) Longitudinal resistivity ($\rho_{yy}$); (d) Hall resistivity ($\rho_{zy}$). The black vertical arrows in all the panels indicate variation of the temperature.

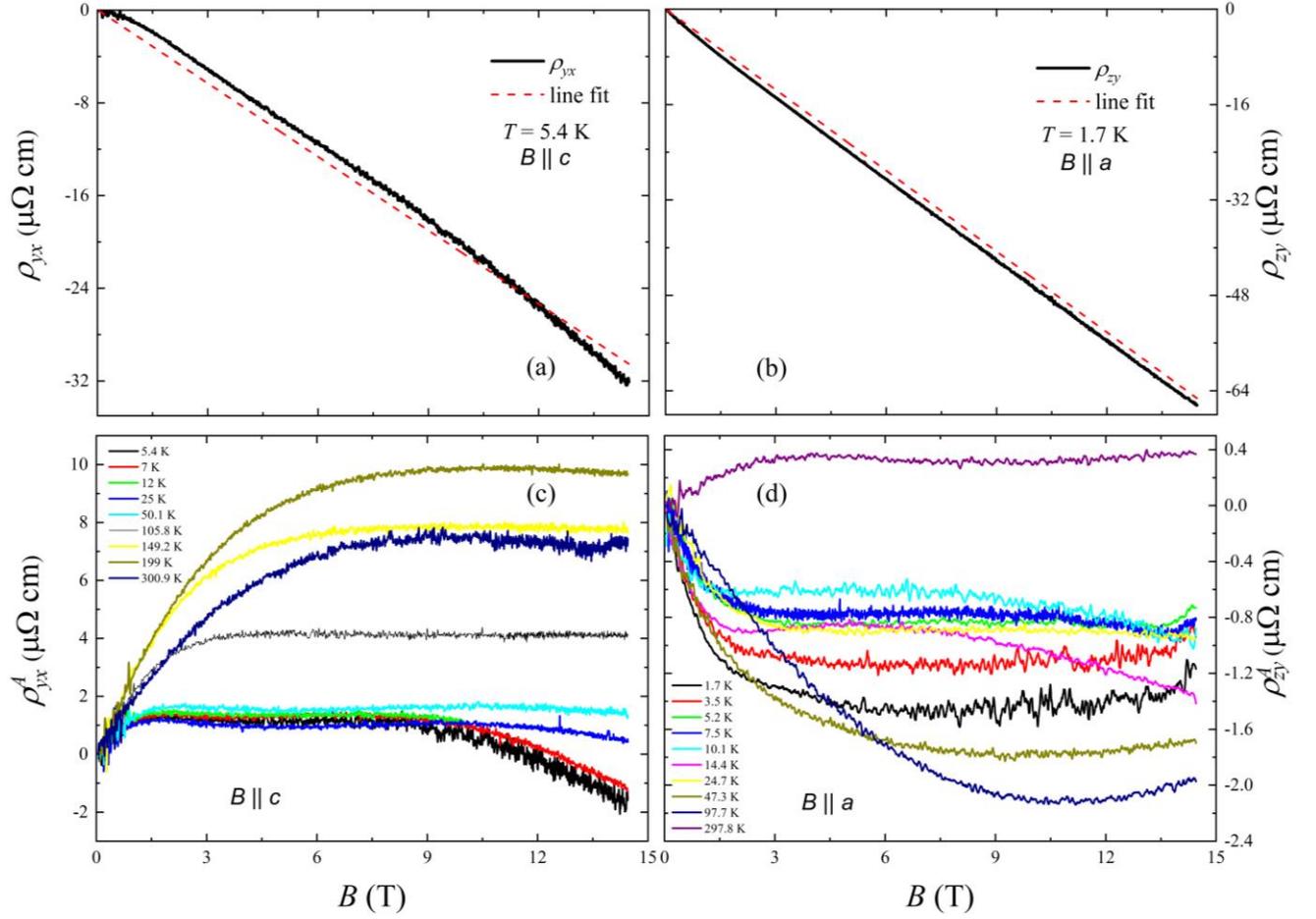

**Figure 3.** (Color online) Magnetic field dependences of the Hall resistivity in CeAlSi: (a) the Hall resistivity ($\rho_{yx}$) as a function of $B$ at $T = 5.4$ K (black line); (b) the Hall resistivity ($\rho_{zy}$) as a function of $B$ at $T = 1.7$ K (black line). Red dashes lines in panels a and b represent the the high field ($B > 3$ T) linear fits. (c) The anomalous contribution to the Hall resistivity ($\rho_{yx}^A$) extracted from $\rho_{yx}(B)$ for several temperatures. (d) The anomalous contribution to the Hall resistivity ($\rho_{zy}^A$) extracted from $\rho_{zy}(B)$ for several temperatures.

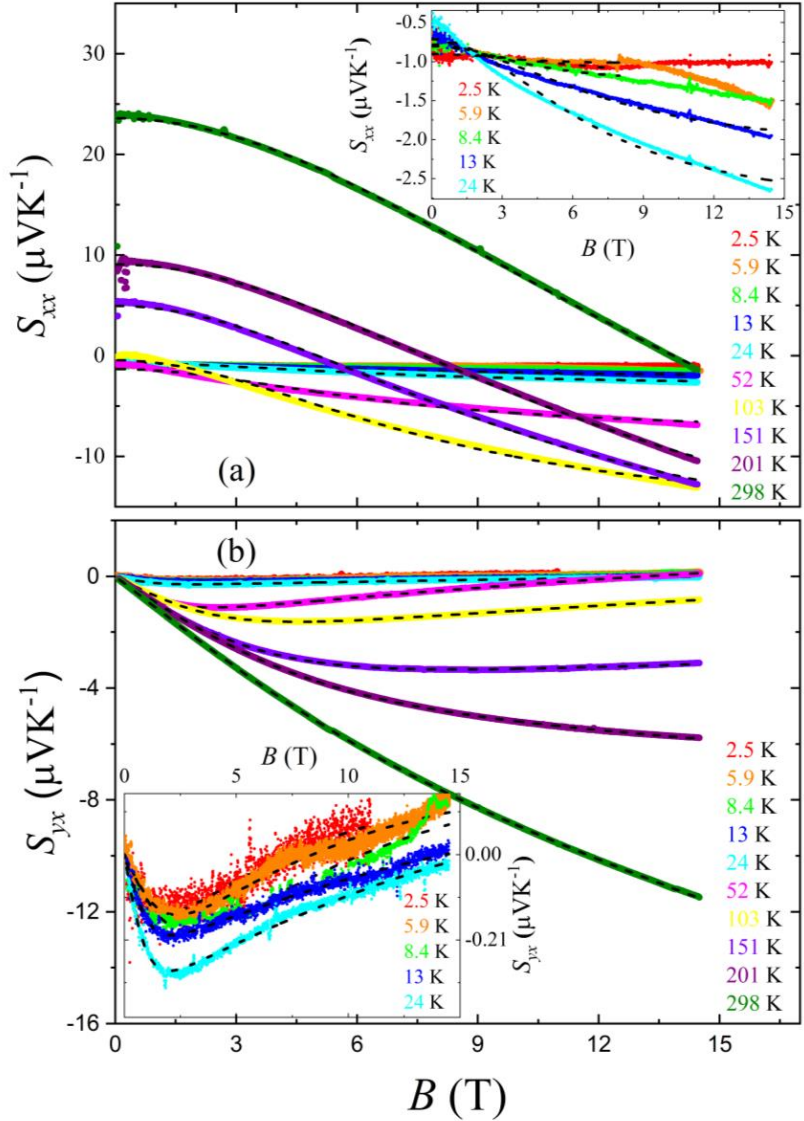

**Figure 4.** (Color online) The magnetic field dependence of the Seebeck (panel a) and Nernst (panel b) signal in CeAlSi for various temperatures. Insets show low temperatures field dependences of the respective coefficients. The black dashes lines in panels (a) and (b) show fits prepared using Eqn. 2 & 4.

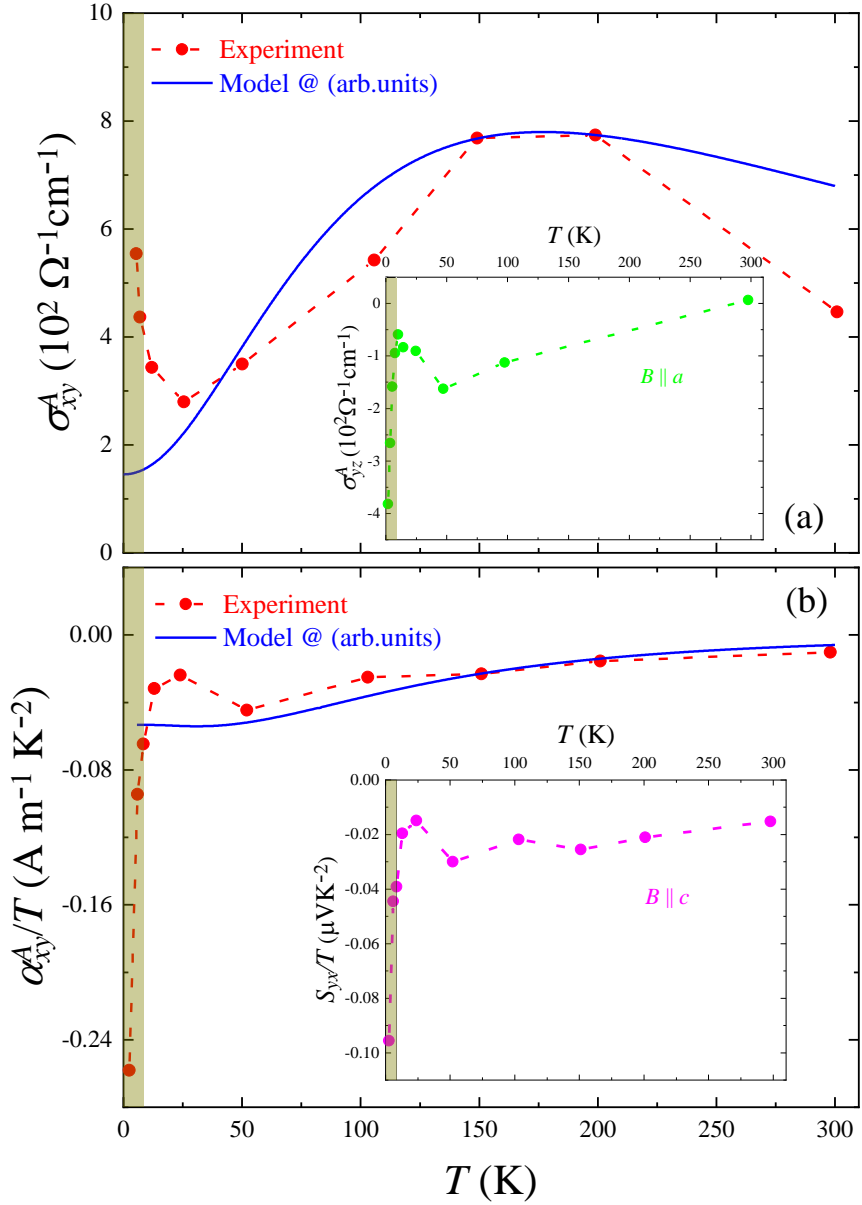

**Figure 5.** (Color online) The temperature dependences of the anomalous Hall conductivity ($\sigma_{xy}^A$) (panel a) and the anomalous Nernst conductivity (panel b) for $B \parallel c$ in CeAlSi. Inset in the upper panel shows temperature dependent anomalous conductivity ($\sigma_{yz}^A$) for $B \parallel a$; inset in the lower panel presents the temperature dependence of normalized anomalous Nernst effect for $B \parallel c$. Blue solid lines in both panels presents the $\sigma_{xy}^A$ and $\alpha_{xy}^A/T$ temperature dependences calculated in arbitrary units using Eqn. 8 & 9. Vertical dark yellow areas in all panels represent the FM regime.

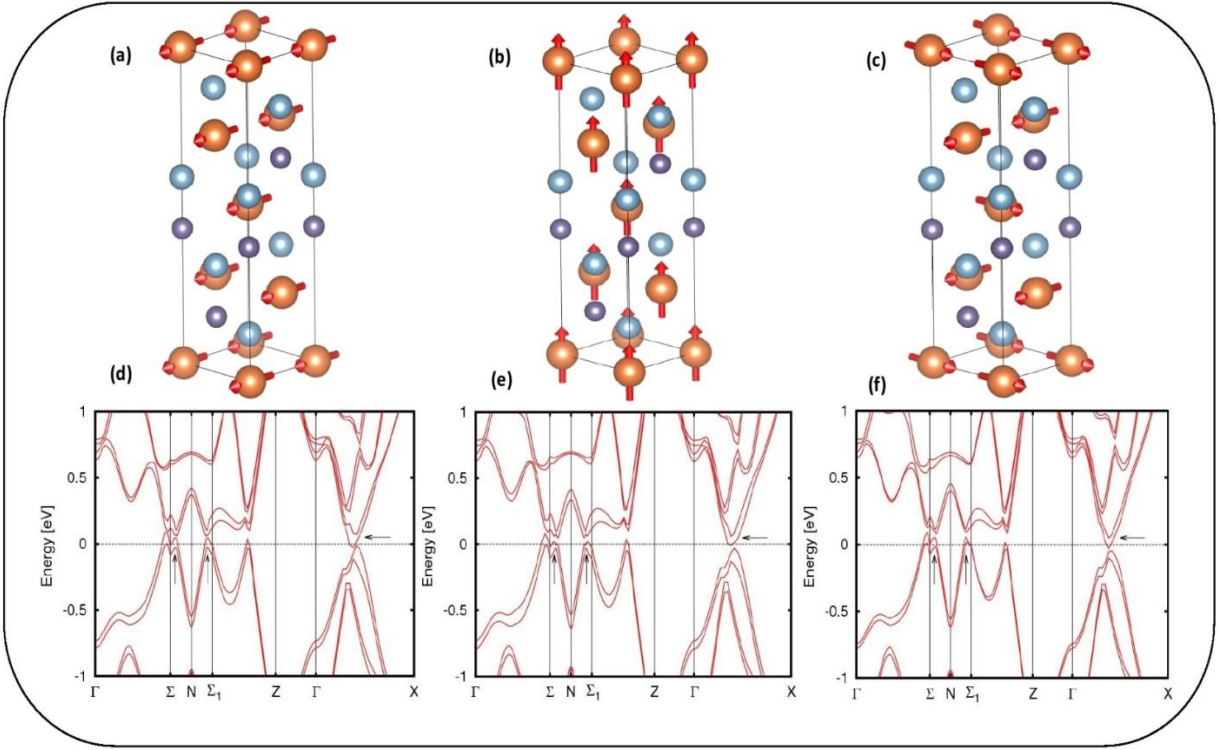

**Figure 6**. Magnetic configurations and associated band structures of the CeAlSi Weyl semimetal. (a) Collinear FM order with spins aligned along the *a*-axis. (b) Collinear FM order with spins aligned along the *c*-axis. (c) Noncollinear FM order. (d), (e) and (f) represent the band structure of CeAlSi along the high symmetry paths including spin-orbit coupling in the three mentioned configurations respectively. The vertical arrows represent the band crossings at the Fermi surfaces between **Σ-N** and N-$\Sigma_i$ , while the horizontal arrow point at the minimum of the conduction band between **Γ** and **X**.

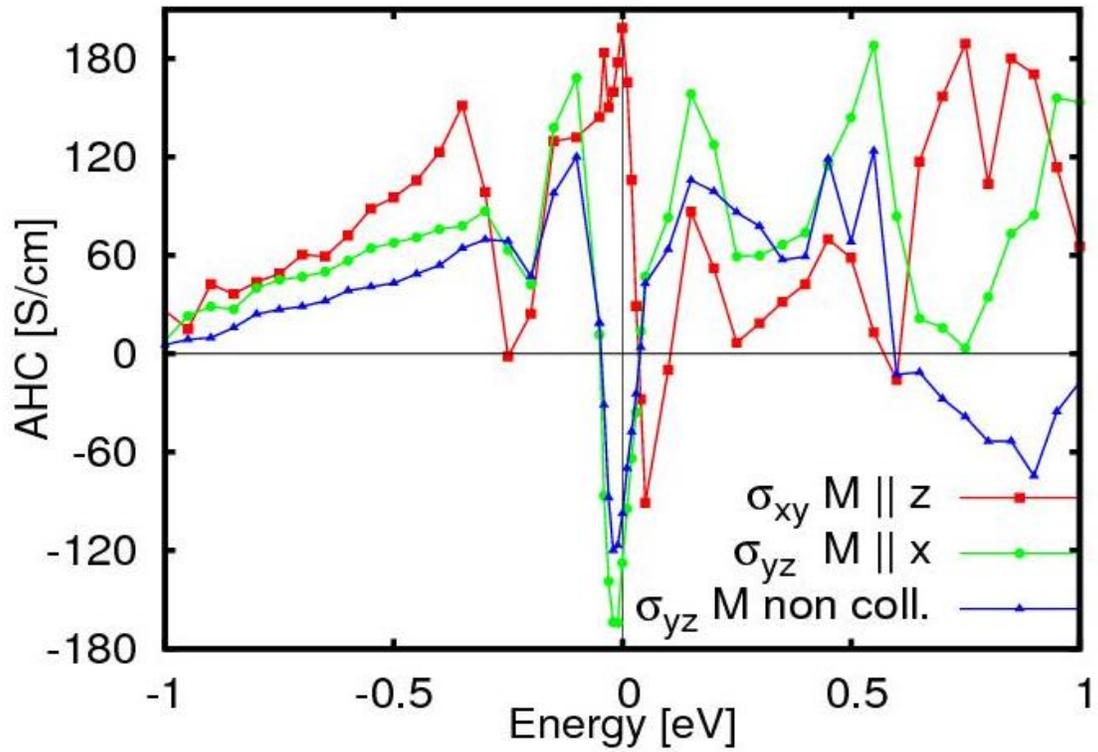

**Figure 7**. Calculated intrinsic anomalous Hall conductivity for the collinear ferromagnetic configuration with spins along the a-axis (green with circle points), along the c-axis (red with square points), and in the noncollinear magnetic configuration (blue with triangle points). The energy range is between -1 and +1 eV. The Fermi level is set to zero for all three magnetic configurations.

Supplemental material:

# Sign change of the anomalous Hall effect and the anomalous Nernst effect in Weyl semimetal CeAlSi

Md Shahin Alam [1,*], Amar Fakhredine [2], Mujeeb Ahmed [1], P.K. Tanwar [1], Hung-Yu Yang [3], Fazel Tafti [3], Giuseppe Cuono [1], Rajibul Islam [1], Bahadur Singh [4], Artem Lynnyk [2], Carmine Autieri [1,†], Marcin Matusiak [1,5,‡]

6. *International Research Centre MagTop, Institute of Physics, Polish Academy of Sciences, Aleja Lotników 32/46, PL-02668 Warsaw, Poland*

7. *Institute of Physics, Polish Academy of Sciences, Aleja Lotników 32/46, PL-02668 Warsaw, Poland*

8. *Department of Physics, Boston College, Chestnut Hill, Massachusetts 02467, USA*

9. *Department of Condensed Matter Physics and Materials Science, Tata Institute of Fundamental Research, Mumbai 400005, India*

10. *Institute of Low Temperature and Structure Research, Polish Academy of Sciences, ul. Okólna 2, 50-422 Wrocław, Poland*

## A: Non-anomalous fits of the magnetic field dependences

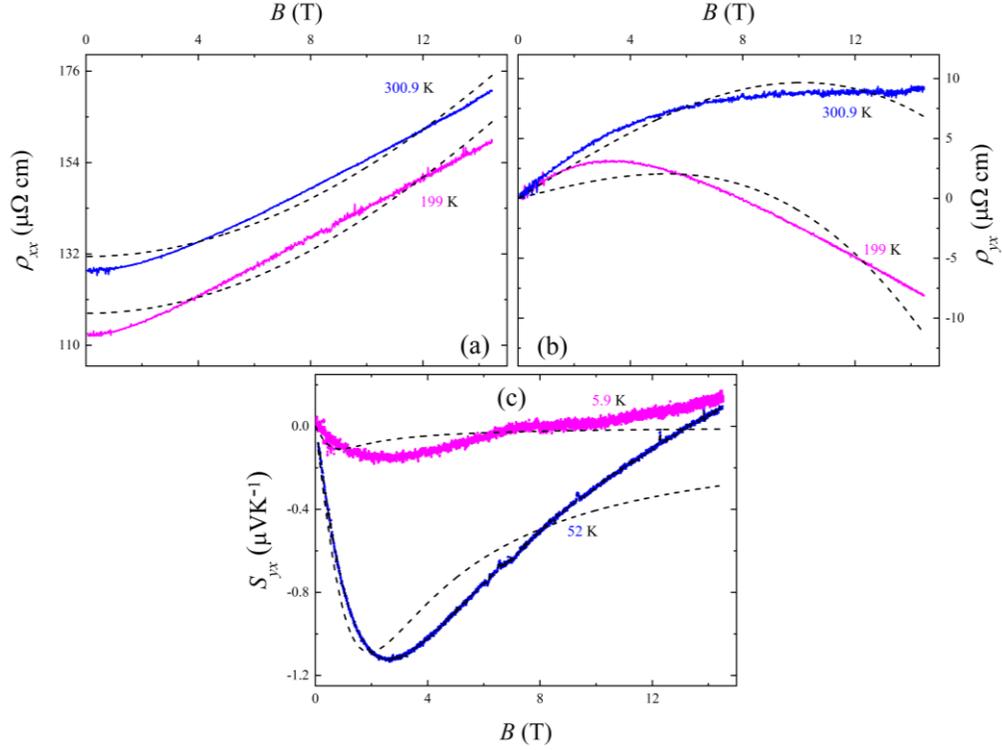

**Figure S1**. (a & b) Magnetic field dependences of the longitudinal resistivity ($\rho_{xx}$) and Hall resistivity ($\rho_{yx}$) for $T$ = 199 K (magenta) and 300.9 K (blue) for CeAlSi. The black dashed lines indicate the best simultaneous fits using the semi-classical two band approximation [1]. There is an apparent discrepancy between the experimental data and the model. (c) Nernst signal ($S_{yx}$) as a function of $B$. Black dashed lines represent fits to: $S_{yx}^N = S_0^N \frac{\mu}{1+(\mu B)^2}$ (see main text), which poorly matches the experimental data.

## B: Scaling of anomalous Hall resistivity

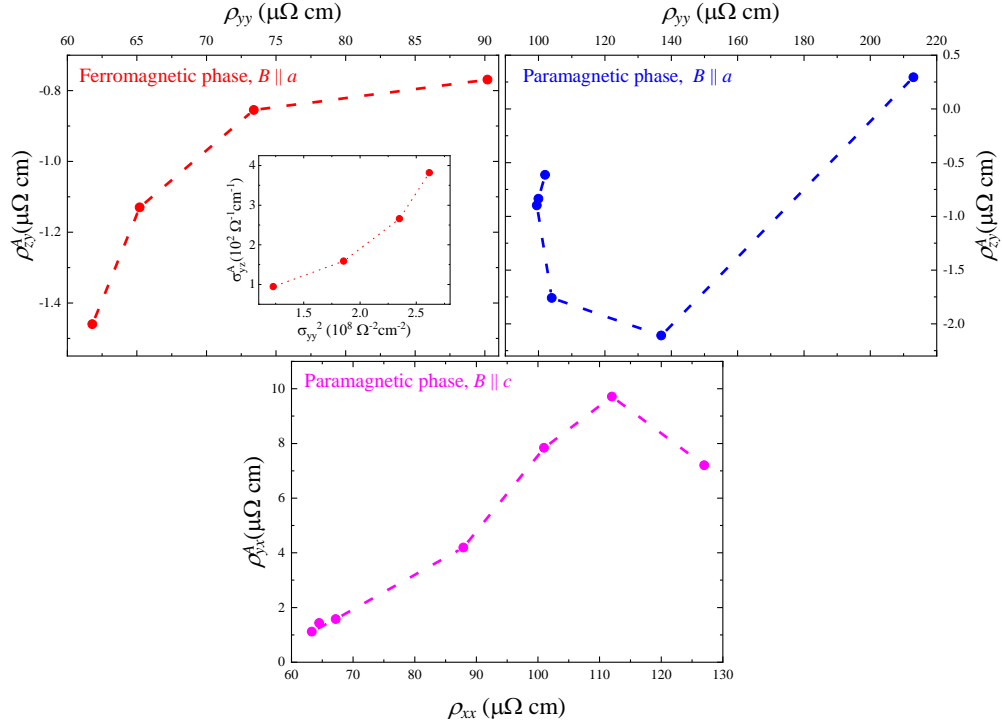

**Figure S2**. (a) The anomalous Hall resistivity ($\rho_{zy}^A$) in ferromagnetic phase of CeAlSi plotted versus longitudinal resistivity ($\rho_{yy}$) for $B \parallel a$. Inset shows the anomalous Hall conductivity ($\sigma_{yz}^A$) as a function of the square of the longitudinal conductivity ($\sigma_{yy}^2$) (b) $\rho_{zy}^A$ vs $\rho_{yy}$ for $B \parallel a$ in paramagnetic phase. (c) The anomalous Hall resistivity ($\rho_{yx}^A$) versus the longitudinal resistivity ($\rho_{xx}$) for $B \parallel c$, in paramagnetic phase.

## C: Magnetic measurements

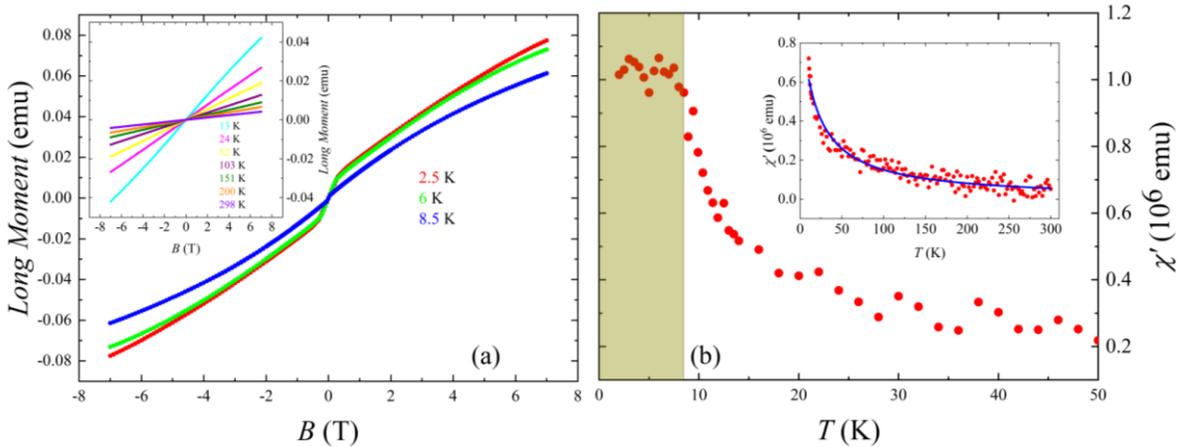

**Figure S3**. (a) Magnetic field dependences of the longitudinal magnetic moments in ferromagnetic phase of CeAlSi. Inset shows long magnetic moment as a function of $B$ in the paramagnetic phase. (b) Temperature dependence of the magnetic susceptibility ($\chi'$) with the ferromagnetic region marked in dark yellow. Inset shows $\chi'$ vs $T$ (red points) in the paramagnetic phase fitted to the Curie-Weiss law (solid blue line).

**D: Anomalous Hall and Nernst conductivities in presence of non-zero Berry curvature**

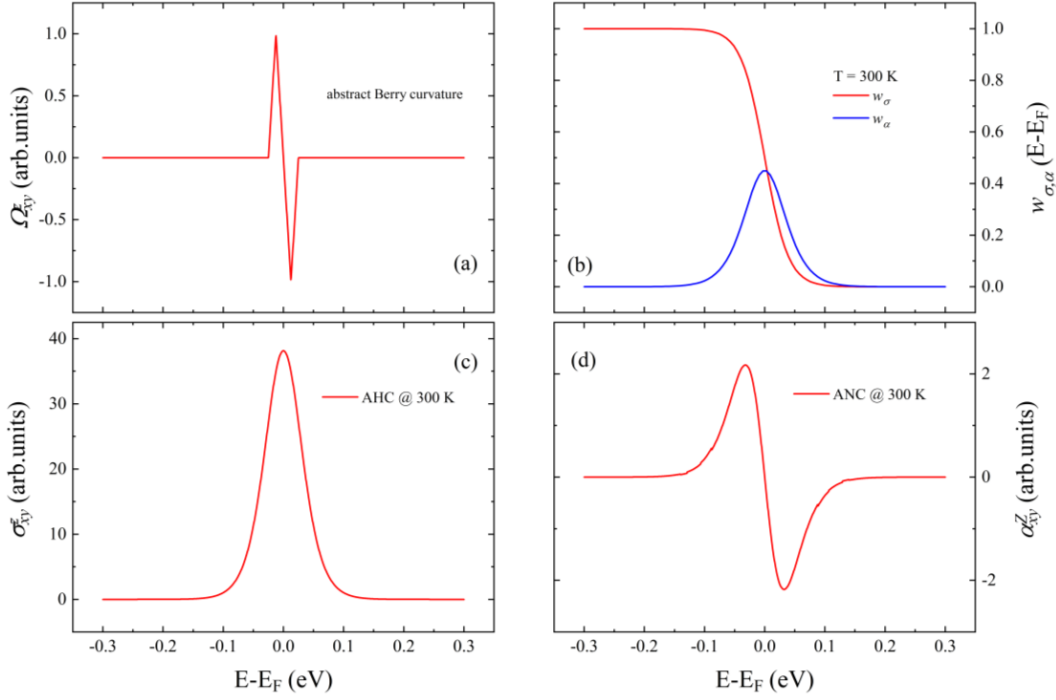

**Figure S4.** (a) Imaginary Berry curvature ($\Omega^z_{xy}$) in arbitrary units plotted versus energy near the Fermi level. (b) Weighting function ($\omega$) for the anomalous Hall conductivity (red) and the transverse thermoelectrical conductivity (blue) as a function of energy at $T = 300$ K. (c) Energy dependences of the anomalous Hall conductivity at $T = 300$ K, calculated as a product of $\Omega^z_{xy}$ and weighting function $w_\sigma$. (d) Anomalous Nernst conductivity as a function of energy $T = 300$ K calculated as a product of $\Omega^z_{xy}$ and weighting function $w_\alpha$.

# E: Computational details for Density functional theory

First-principles calculations were conducted and the projector-augmented-wave (PAW) method was applied. The generalized gradient approximation of the Perdew-Burke-Ernzerhof(PBE) form was used to treat the exchange-correlation energy (GGA), as implemented in the Vienna ab-initio simulation package (VASP) in the frame of the density functional theory [2,3]. The cut-off energies for the plane-wave basis set that were used to expand the Kohn-Sham orbitals were chosen to be 450 eV. All calculations include the spin-orbit coupling. The Monkhorst-Pack scheme was used for the Γ-centered 10×10×10 k-point sampling. The calculations are performed using the experimental lattice parameters of CeAlsi (a = 4.252 Å; c = 14.5801 Å) [4]. In order to conduct the effect of strongly correlated electrons in the 4f-shell of Ce, we applied the GGA+U method within the self-consistent DFT cycle [5]. The on-site Coulomb interaction strength U applied to the Ce 4f states is 7 eV and the intra-atomic exchange interaction strength $J_H$ was taken as zero. Our ab-initio results were utilized to obtain the Wannier tight-binding Hamiltonian using the VASP2WANNIER90 interface [6]. The Anomalous Hall Effect was calculated subsequently after applying the technique of Wannier interpolation [7] using a k-grid of 100 x 100 x 100 with an adaptive grid of 5 x 5 x 5. The AHE calculations were verified with a k-grid of 300 x 300 x 300 that reproduce the same results with negligible differences. The calculations were performed on three different spin configurations of the CeAlSi unit cell; a collinear FM phase with the spins aligned along the x-axis direction, a collinear FM phase with the spins aligned along the z-axis direction, and a non-collinear FM ordering. All the AHE calculations reproduce a minimum and a maximum close to the Fermi level, therefore the AHE results are robust respect to the wannierization of the three different magnetic configurations.

For better analysis of the electronic structure of CeAlSi, the atom-resolved local density of states (LDOS) was also calculated for the non-collinear ferromagnetic configuration with spin-orbit coupling (SOC) and presented in Figure 8. DOS for other magnetic configurations are qualitatively similar. The sp-bands of Al and Si are present both below and above the Fermi level. One band of the Ce 4f-states is present at -3 eV from the Fermi level, while the other Ce 4f-states are above 2 eV from the Fermi level. The sp-orbitals of Ce are less relevant, while close to the Fermi level, the only relevant contribution from the Ce atoms comes from the 5d-states.

The oxidation states of atoms is zero in this compound. The electronic configuration for Ce reported in textbooks is [Xe]$4f^1 5d^1 6s^2$ in the ionic approximation, this electronic configuration can change depending on the structural phases and crystal fields. In CeAlsi, the electronic configuration changes to be something more similar to $4f^1 5d^3 6s^0$ in the ionic approximation according to our local DOS. Indeed, due to the crystal field effect of the surrounding atoms that act differently on d-electrons and on s-electrons, the s-electrons of Ce are pushed to higher energies. Since we are only interested in the electronic properties at the Fermi level, we have approximated the Ce atom as [Xe]$4f^1 5d^3 6s^0$ considering that only the 5d electrons are relevant for the wannierization and excluding the 4f electrons from our tight-binding model. For this reason, the wannierization was performed taking into consideration the sp-electrons of Al , sp-electrons of Si and the d-electrons of Ce as tight-binding basis. Within its simplicity, this approximation brings us to a quite good result of the wannierization. We have chosen the frozen window around the Fermi level excluding the 4f states in both conduction and valence band. Our wannierization is in agreement with that reported in the literature [8].

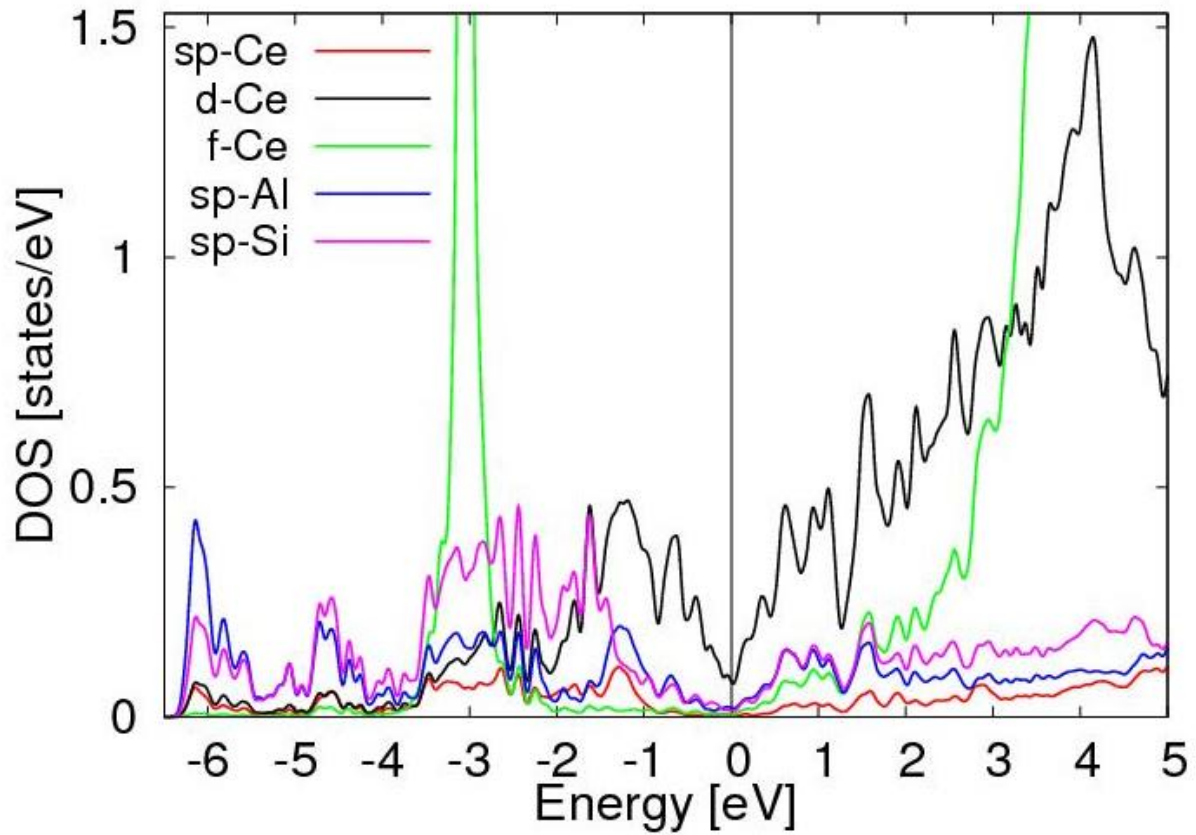

**Figure S5**. Local density of states for the non-collinear magnetic configuration of CeAlSi. The 4f-levels of Ce are extremely localized and their DOS goes beyond the maximum of the y-range. The Fermi level is set to zero.